\documentclass[11pt,USenglish]{article}
\usepackage{amsmath}
\usepackage{amssymb}
\usepackage{amsthm}
\usepackage[english]{babel}
\usepackage{fullpage}
\usepackage{graphicx}
\usepackage[colorlinks=true,linkcolor=black,citecolor=black]{hyperref}
\usepackage{xcolor}

\usepackage{color}

\definecolor{darkgreen}{rgb}{0,.5,.2}

{\vspace{5mm}\begin{sloppypar}
  \noindent\hrulefill\, BEGIN \,\hrulefill
  \end{sloppypar}
}%
{\begin{sloppypar}
  \noindent\hrulefill\, END   \,\hrulefill
  \end{sloppypar}\vspace{5mm}
}

\newcounter{todo}

\newcounter{question}

\newcounter{comment}

\theoremstyle{plain}

\newtheorem{example}{Example}

\theoremstyle{definition}
\newtheorem{remark}{Remark}

\newtheoremstyle{mydefinition}
	{4pt} 
	{-8pt} 
	{} 
	{} 
	{\bfseries} 
	{.} 
	{.5em} 
	{} 

\theoremstyle{mydefinition}
\newtheorem{defn}{Definition}
\newenvironment{definition}%
	{\textcolor{black!30}{\hrule}\begin{defn}}%
	{\end{defn}\textcolor{black!30}{\hrule}\bigskip}

\newcommand{\definedTerm}[1]{\textbf{#1}}

\newcommand{\mathSmallRDF}[1]{{\normalfont\texttt{\small #1}}}

\newcommand{\definedAs}    
	{=}

\newcommand{\fctsymDom}{\mathrm{dom}} 
\newcommand{\fctDom}[1]{\fctsymDom(#1)}

\newcommand{\tuple}[1]{\langle #1 \rangle} 
\newcommand{\bigtuple}[1]{\big\langle #1 \big\rangle} 

\newcommand{\symAllIRIs}{\mathcal{I}} 
\newcommand{\symAllLiterals}{\mathcal{L}} 
\newcommand{\symAllBNodes}{\mathcal{B}} 
\newcommand{\symAllTriples}{\mathcal{T}} 
\newcommand{\symIRI}{u} 

\newcommand{\symAllStarTriples}{\mathcal{T}^{\!\star\!}} 
\newcommand{\symAllStarTriplesInfinite}{\widetilde{\symAllTriples}} 

\newcommand{\symRDFstarGraph}{G^{\star}} 

\newcommand{\fctsymTermsPlus}{\mathrm{Elmts\!}^+} 
\newcommand{\fctTermsPlus}[1]{\fctsymTermsPlus\!(#1)} 

\newcommand{\fctsymTRefs}{\mathrm{Emb\!}^+} 
\newcommand{\fctTRefs}[1]{\fctsymTRefs\!(#1)} 

\newcommand{\symPGraph}{G}
\newcommand{\symAllDataTypes}{\mathfrak{D}} 
\newcommand{\symAllDataValues}{\mathcal{V}} 
\newcommand{\symDataType}{\mathcal{D}} 
\newcommand{\symDataTypeString}{\mathcal{S}} 
\newcommand{\str}[1]{\text{\normalfont\texttt{"#1"}}}

\newcommand{\xxxSTAR}
	{$^\star$}

\newcommand{\rdfStar}{RDF\xxxSTAR}
\newcommand{\turtleStar}{Turtle\xxxSTAR}
\newcommand{\sparqlStar}{SPARQL\!\xxxSTAR}

\newcommand{\rdfStarTriple}{{\rdfStar}\! triple}
\newcommand{\rdfStarGraph}{{\rdfStar}\! graph}

\newcommand{\tRef}%
	{tref}

\title{Reconciliation of \rdfStar\ and Property Graphs}
\author{Olaf Hartig \\ {\normalsize University of Waterloo} \vspace{1mm} \\ {\normalsize http://olafhartig.de}}

\begin{document}
\maketitle

\vspace*{-5mm} 

\begin{abstract}
	\noindent
Both the notion of Property Graphs (PG) and the Resource Description Framework (RDF) are commonly used models for representing graph-shaped data.
While there exist some sys\-tem-spe\-cif\-ic solutions to convert data from one model to the other, these solutions are not entirely compatible with one another and none of them appears to be based on a formal foundation.
	In fact, for the PG model, there does not even exist a commonly agreed-upon formal definition.

	The aim of this document is to reconcile both models formally. To this end, the document proposes a formalization of the PG model and introduces well-de\-fined transformations between PGs and RDF.
As a result, the document provides a basis for the following two innovations: On one hand, by implementing the RDF-to-PG transformations defined in this document, PG-based systems can enable their users to load RDF data and make it accessible in a \emph{compatible}, sys\-tem-in\-de\-pen\-dent manner using, e.g., the graph traversal language Gremlin or the declarative
	graph
query language Cypher.
On the other hand, the PG-to-RDF transformation in this document enables RDF data management systems to support \emph{compatible}, sys\-tem-in\-de\-pen\-dent queries over the content of Property Graphs by using the standard RDF query language
	SPARQL.
Additionally, this document represents a foundation for
	systematic research on relationships between the two models and between their query languages.

\bigskip
\noindent
\textbf{If you have comments or suggestions about this proposal,
	as well as implementations or applications
%
	thereof,
do not hesitate to let the author know about them.}
\end{abstract}

\section{Introduction} \label{sec:Intro}
	This document reconciles
two commonly used graph-based data models, namely the Property Graphs model which is used by popular graph database systems such as Neo4j\footnote{\url{http://neo4j.org/}}\!, Titan\footnote{\url{http://thinkaurelius.github.com/titan/}} and Sparksee\footnote{\url{http://sparsity-technologies.com/\#sparksee}}\!, and the RDF data model~\cite{Cyganiak14:RDFConcepts}, which has been standardized by the World Wide Web Consortium (W3C) and is supported by numerous systems including IBM's DB2\footnote{\url{http://www-01.ibm.com/software/data/db2/linux-unix-windows/nosql-support.html}}\!, OpenLink's Virtuoso\footnote{\url{http://virtuoso.openlinksw.com/}} and Systap's Bigdata\footnote{\url{http://www.systap.com/}}\!. The
	primary
goal of this reconciliation is to enable a user who is familiar with either of these data models to access data, represented in the other model,
	based on a \emph{well-de\-fined, sys\-tem-in\-de\-pen\-dent view} of this data given in the model most familiar to the~user.

The
	document is structured as follows:
Section~\ref{ssec:Intro:Models} gives a very brief informal overview of
	Property Graphs, RDF, and an extension of RDF that provides for a more us\-er-friend\-ly representation of state\-ment-lev\-el metadata and serves as a basis of the reconciliation.
Section~\ref{ssec:Intro:Proposal} then describes the proposal informally.
Thereafter, Sections \ref{sec:Preliminaries} to~\ref{sec:TransformPGtoRDFstar} provide the formal details.

\section{Informal Overview of the Data Models} \label{ssec:Intro:Models}

The notion of a Property Graph has been described informally in a number of books, tutorials, manuals, and other documentation on the Web. Perhaps one of the most representative descriptions is the following by Robinson et al.~\cite{Robinson13:GraphDBsBook}.
\begin{quote}
\itshape
\begin{itemize}
	\item
		``A property graph is made up of nodes, relationships, and properties.
	\item
		Nodes contain
			properties [...]
		in the form of arbitrary key-value pairs. The keys are strings and the values are arbitrary data types.
	\item
			$\text{[}$...]
		A relationship always has a direction, a label, and a start node and an end
			node [...].
	\item
		Like nodes, relationships can also have
			properties.''~\cite{Robinson13:GraphDBsBook}
\end{itemize}
\end{quote}

\begin{example} \label{ex:PGinformal}
	Figure~\ref{fig:ExamplePG} illustrates a simple Property Graph with two nodes and two relationships between them. One of these relationships has the label \textsl{mentioned}, the start node
	is the node for Orson Welles, and the end node is the node for Stanley Kubrick. The other relationship, labeled \textsl{influencedBy}, starts from the Kubrick node and ends in the Welles node.
	The light gray boxes associated with some of the graph elements represent properties of these elements. For instance, the node for Stanley Kubrick has two properties giving the name and the birth year of the famous director, and the \textsl{influencedBy} relationship has a property that specifies the certainty of whether Kubrick was influenced by Welles. The other relationship does not have a property in the given~graph.
\begin{figure}[ht]
	\centering
	\includegraphics[width=0.8\textwidth]{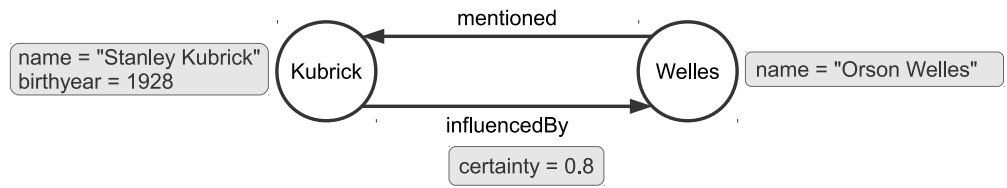}
	\caption{A Property Graph with two nodes.}
	\label{fig:ExamplePG}
\end{figure}
\end{example}

\noindent
In contrast to Property Graphs, RDF is a standardized data model. This model represents data as sets of triples where each triple consists of three elements that are referred to as the subject,~the predicate, and the object of the triple.
	These triples allow
users to describe arbitrary things in~terms of their attributes and their relationships to other things. That is, such things may be the subject~or object of an RDF triple in which, usually, they are denoted by an Internationalized Resource Identifier (IRI)---which is a form of a Web-wide unique identifier (hence, in contrast to vertex and edge identity that exists only within a given Property Graph, IRIs identify things across datasets). Relationships and attributes are also denoted by IRIs, which appear as the predicate of RDF triples. Finally, attribute values in RDF are called literals and appear in the object of RDF triples.
	Any set of RDF triples can be conceived of as a directed graph in which each triple represents an edge from the subject to the object of that triple; hence, the vertices in this graph are all subjects and objects of all triples in~the~set.

\begin{figure}[t]
	\centering
	\includegraphics[width=0.7\textwidth]{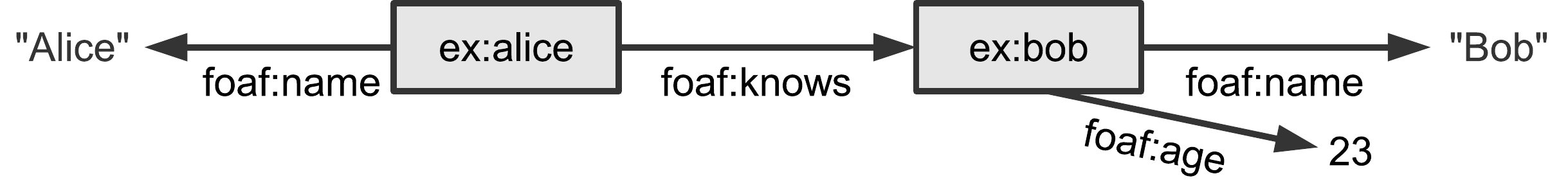}
	\caption{A simple RDF graph consisting of four RDF triples.}
	\label{fig:ExampleRDFgraph}
\end{figure}

\begin{example} \label{ex:RDFinformal}
	The following RDF data, represented in the Turtle format~\cite{PrudHommeaux14:Turtle}, contains four RDF triples that describe two persons, Alice and Bob,
	denoted by the IRIs \texttt{http://example.org/alice} and \texttt{http://example.org/bob}, respectively. The objects of the last three triples are literals. Figure~\ref{fig:ExampleRDFgraph} illustrates the graph representation of these four triples.
\begin{small}
	\begin{verbatim}
	  @prefix foaf: <http://xmlns.com/foaf/0.1/> .
	  @prefix ex: <http://example.org/> .

	  ex:alice  foaf:knows  ex:bob .
	  ex:alice  foaf:name   "Alice" .
	  ex:bob  foaf:name   "Bob" .
	  ex:bob  foaf:age    23 .
	\end{verbatim}
\end{small}
\end{example}


\noindent
A
	shortcoming
that RDF has been widely criticized for is
	the lack of an approach to represent state\-ment-lev\-el metadata that is as intuitive and us\-er-friend\-ly as edge properties in Property Graphs~(such as the certainty statement in the sample Property Graph in Figure~\ref{fig:ExamplePG}).
While
	RDF provides a notion of reification to support this use case~\cite{Hayes14:RDFSemantics},
this approach is awkward to use, the resulting metadata is cumbersome to query, and it may blow up the dataset size significantly.
However, a recently proposed extension of RDF addresses this shortcoming by making \emph{triples about triples} a first class citizen in the data model~\cite{Hartig14:RDFStar}. That is, this
	extension, which is called \rdfStar\!, augments RDF with the possibility to use a triple directly as the subject or object of other triples%
	.

\begin{example} \label{ex:RDFstarInformal}
	Consider an
		extended version of the RDF Turtle format, called \turtleStar~\cite{Hartig14:RDFStar}, 
	that implements the idea of embedding RDF triples into other RDF triples by enclosing the embedded triple in '{\scriptsize\textless\textless}' and '{\scriptsize\textgreater\textgreater}'. Then, the data in Example~\ref{ex:RDFinformal} can be augmented with
		metadata as follows:
\begin{small}
	\begin{verbatim}
	  <<ex:alice foaf:knows ex:bob>>  ex:certainty  0.5 .
	  ex:alice  foaf:name   "Alice" .
	  ex:bob  foaf:name   "Bob" .
	  <<ex:bob foaf:age 23>>  ex:certainty  0.9 .
	\end{verbatim}
\end{small}
	Note that the last line represents a metadata triple whose subject is the triple that provides information about the age of Bob, and the metadata triple describes the certainty of this information. Similarly, the first line represents a metadata triple about the triple that is embedded as subject.
\end{example}

\noindent
It is important to emphasize that \rdfStar\ is simply a syntactic extension of RDF that makes dealing with state\-ment-lev\-el metadata more intuitive. In fact, there exists a well-de\-fined transformation%
	\footnote{In a nutshell, this transformation employs standard RDF reification to convert any \rdfStar\ metadata triple into a set of ordinary RDF triples~\cite{Hartig14:RDFStar}.}
of \rdfStar\! data back to standard RDF data~\cite{Hartig14:RDFStar}.
On the other hand, standard RDF data can be understood trivially as \rdfStar\! data. 
	This document leverages these properties and
%
	uses \rdfStar\ as a basis for reconciling the RDF data model and the Property Graphs model.

\section{Informal Overview of the Proposal} \label{ssec:Intro:Proposal}
This document \emph{formalizes} three transformations: two
from \rdfStar\! data to Property Graphs, and one from Property Graphs to \rdfStar\! data.
	All three transformations also cover ordinary RDF data because of the aforementioned relationship between \rdfStar\ and the standard RDF data model.
This section provides an informal overview of the transformations and outlines some use cases.

\subsection{First Transformation: \rdfStar\ to RDF-like Property Graphs} \label{ssec:Informal:TransformRDFstarToRDFlikePGs}
	The
first transformation
	presents an intuitive (perhaps the most natural) way of converting \rdfStar\! data to Property Graphs; namely, this transformation
represents any ordinary RDF triple as an edge in the resulting Property Graph;
	the two vertices incident to such an edge have properties that describe the subject and the object of the corresponding RDF triple;
and metadata triples are represented as edge properties.

\begin{example} \label{ex:InformalResultOf1stTransformation}
	Figure~\ref{fig:ExampleResultingRDFlikePG} illustrates the Property Graph that can be obtained by transforming the \rdfStar\! data in Example~\ref{ex:RDFstarInformal} based on the proposed transformation.
%
\begin{figure}[ht]
	\centering
	\includegraphics[width=0.9\textwidth]{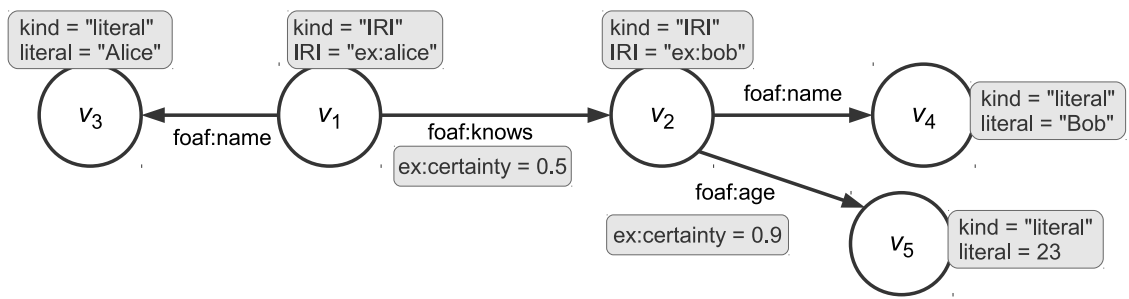}
	\caption{An (RDF-like) Property Graph that represents the \rdfStar\! data in Example~\ref{ex:RDFstarInformal}.}
	\label{fig:ExampleResultingRDFlikePG}
\end{figure}
\end{example}

\noindent
	A caveat of the straightforward approach to represent \rdfStar\! data as Property Graphs is that the transformation is not applicable in some cases because of
differences in the expressiveness of \rdfStar\ and RDF on the one hand and Property Graphs on the other.
	For instance, the object of a metadata triple (with an RDF triple as its subject) may be an IRI that is also mentioned in other triples, whereas the value of an edge (or vertex) property cannot be a vertex itself.
For a more detailed discussion of these differences and a formal characterization of the \rdfStar\! data that the transformation can be applied to, refer to Section~\ref{ssec:Preliminaries:Convertibility}. A formalization of the transformation itself can be found in~Section~\ref{sec:TransformRDFstarToRDFlikePGs}.

	Even if not generally applicable, for the large number of cases in which the transformation~can be applied, the 
transformation can be used to enable users who are familiar with
the graph query language Cypher to query RDF/\rdfStar\! data without having to learn
	the RDF query language~SPARQL.

\begin{example} \label{ex:QueryingResultOf2ndTransformation}
	The following Cypher query can be used to query the sample data to find the name of persons known by somebody whose name is Alice%
		~(an execution over the sample data would return a single value, namely, Bob)%
	.
\begin{small}
	\begin{verbatim}
	    START a=node(*)
	    MATCH (a)-[:foaf:name]->(bn { literal="Alice" }),
	          (a)-[:foaf:knows]->(p)-[:foaf:name]->(pn)
	    RETURN pn.literal
	\end{verbatim}
\end{small}
	An equivalent\,\footnote{The equivalence cannot be shown formally because Cypher does not have a formally defined query semantics.} SPARQL representation of this Cypher query is given as follows.
\begin{small}
	\begin{verbatim}
	    SELECT ?pn WHERE {
	        ?a  foaf:name  "Alice" .
	        ?a  foaf:knows  ?p .
	        ?p  foaf:name  ?pn }
	\end{verbatim}
\end{small}
%
\end{example}

\noindent
The example demonstrates that Property Graphs obtained by the proposed transformation can be queried using Cypher queries for which there exist corresponding SPARQL
over the original RDF~(or \rdfStar) data. However, the transformation also enables users to benefit from features of Cypher that are not available in SPARQL. For instance, Cypher allows for path expressions that are more powerful than the property path feature provided by SPARQL~1.1.
	Another example is querying state\-ment-lev\-el metadata by accessing the corresponding edge properties using Cypher.

\begin{example} \label{ex:CypherForEdgeProperties}
	Consider an extension of the query in the previous example with the additional condition that the certainty of the information that Alice knows the listed persons must be greater~than~	0.7.
	The following Cypher query captures this condition by filtering on the corresponding edge~property.
\begin{small}
	\begin{verbatim}
	    START a=node(*)
	    MATCH (a)-[:foaf:name]->(bn { literal="Alice" }),
	          (a)-[r:foaf:knows]->(p)-[:foaf:name]->(pn)
	    WHERE r.ex:certainty > 0.7
	    RETURN pn.literal
	\end{verbatim}
\end{small}
\end{example}

\noindent
Any system that is based on Property Graphs can enable users to query RDF (or \rdfStar) data using Cypher as outlined in the examples. To this end, the system only has to implement an import procedure that applies the first (or the second) transformation as formalized in this document. On the other hand, any RDF-based system---independent of whether it
	supports
the \rdfStar\ extension or not---may support Cypher queries on top of a virtual Property Graph view of the data (where the view is defined by the given transformation). However, for \rdfStar\!-en\-abled systems, if the primary use case for supporting Cypher are more us\-er-friend\-ly queries over state\-ment-lev\-el metadata (as demonstrated in Example~\ref{ex:CypherForEdgeProperties}), a more native approach is \sparqlStar, which is an \rdfStar\!-spe\-cif\-ic extension of SPARQL that carries over the idea of embedding triples and comes with a well-de\-fined query~semantics~\cite{Hartig14:RDFStar}.

\begin{example} \label{ex:FirstSPARQLStarQueries}
	The following \sparqlStar\ query is equivalent to the Cypher query in Example~\ref{ex:CypherForEdgeProperties}.
\begin{small}
	\begin{verbatim}
	    SELECT ?pn WHERE {
	        ?a foaf:name "Alice" .
	        <<?a foaf:knows ?p>> ex:certainty ?c .
	        FILTER ( ?c > 0.7 )
	        ?p foaf:name ?pn }
	\end{verbatim}
\end{small}
	An alternative form of this \sparqlStar\ query is to use a BIND clause as follows.
\begin{small}
	\begin{verbatim}
	    SELECT ?pn WHERE {
	        ?a foaf:name "Alice" .
	        BIND( <<?a foaf:knows ?p>>  AS ?t )
	        ?t ex:certainty ?c .
	        FILTER ( ?c > 0.7 )
	        ?p foaf:name ?pn }
	\end{verbatim}
\end{small}
\end{example}

\subsection{Second Transformation: \rdfStar\ to Simple Property Graphs} \label{ssec:Informal:TransformRDFstarToSimplePGs}

Property Graphs as obtained by the transformation described in the previous section resemble the structure of RDF/\rdfStar\! data. In fact, any of these Property Graphs contains all information present in the original \rdfStar\! data. Hence, the transformation is lossless. However, some users
	of Property Graph systems
may consider the resulting Property Graphs unnatural or too complex. Therefore, this document introduces
	another
transformation for converting \rdfStar\! data to Property Graphs that have a simpler structure, but cannot be used to reconstruct the original \rdfStar\! data.

The transformation distinguishes \emph{attribute triples}, that is, ordinary~(non-metadata) triples whose object is a literal, and \emph{relationship triples}, that is, ordinary triples whose object is an IRI or a blank node. Then, the transformation represents every relationship triple as an edge; every attribute triple is converted into a property of the vertex for the subject of that triple~(instead of also converting it into a separate edge as the lossless transformation does). Hence, vertices in the resulting Property Graph represent IRIs and blank nodes only~(whereas the lossless transformation produces vertices that may represent literals). Metadata triples about relationship triples are converted into edge properties, whereas metadata triples about attribute triples cannot be converted by the transformation because the Property Graph model does not support metadata about (vertex) properties. Consequently, the
second transformation
	is more limited in the \rdfStar\! data that it can handle than the
		the aforementioned lossless transformation.

\begin{example} \label{ex:InformalResultOf3rdTransformation}
	The first line in the \rdfStar\! data in Example~\ref{ex:RDFstarInformal} represents a metadata triple about a relationship triple, and the fourth line represents represents a metadata triple about an attribute triple. Due to the existence of the latter metadata triple, the lossy transformation cannot be applied to the given sample data.
	However, suppose the sample data would not contain the last line. Then, Figure~\ref{fig:ExampleResultingSimplePG} illustrates the (simple) Property Graph that can be obtained by transforming this data based on the lossy transformation.
	A Cypher query for this Property Graph with the same intention as the queries in Examples \ref{ex:CypherForEdgeProperties} and~\ref{ex:FirstSPARQLStarQueries} is given as follows.
\begin{small}
	\begin{verbatim}
	    START a=node(*)
	    MATCH (a { foaf:name="Alice" })-[r:foaf:knows]->(p)
	    WHERE r.ex:certainty > 0.7
	    RETURN p.foaf:name
	\end{verbatim}
\end{small}

\begin{figure}[t]
	\centering
	\includegraphics[width=0.65\textwidth]{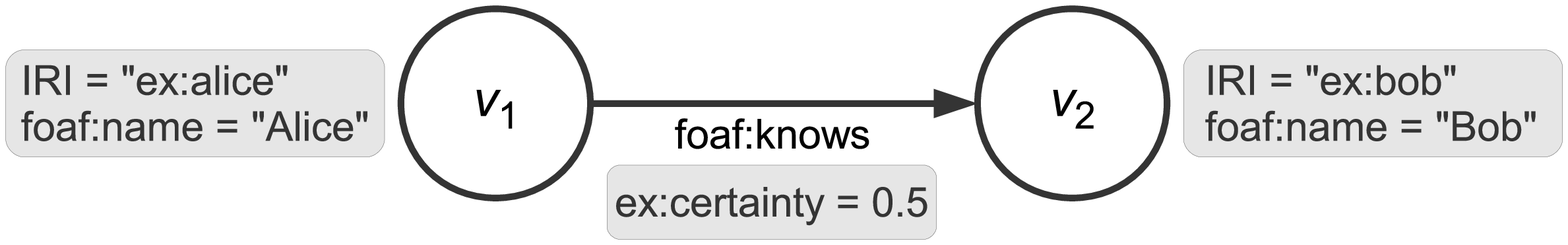}
	\caption{A simple Property Graph that represents
		the \rdfStar\! data in Example~\ref{ex:RDFstarInformal} partially.}
	\label{fig:ExampleResultingSimplePG}
\end{figure}
\end{example}

\subsection{Third Transformation: Property Graphs to \rdfStar} \label{ssec:Informal:TransformPGtoRDFstar}

The third transformation
	converts
Property Graphs to \rdfStar\! data~(which then may be transformed to standard RDF data~\cite{Hartig14:RDFStar}). The idea of the transformation is simple: Every edge
	(including its label)
in a given Property Graph
is represented as an ordinary RDF triple in the resulting \rdfStar\! data; the same holds for every vertex property. Any edge property is represented as a metadata triple whose subject is the triple representing the corresponding edge.
The transformation gives users the freedom to choose patterns for generating IRIs that denote edge labels and properties keys, respectively. These IRIs become the predicates of triples in the resulting \rdfStar\! data.

\begin{example} \label{ex:InformalResultOf2ndTransformation}
	An \rdfStar\ representation (given in \turtleStar\ format)
		that is the result of applying the proposed transformation to
	the Property Graph in Example~\ref{ex:PGinformal} (cf.~Figure~\ref{fig:ExamplePG}) is given as follows:
\begin{small}
	\begin{verbatim}
	  @prefix p: <http://example.org/property/> .
	  @prefix r: <http://example.org/relationship/> .

	  _:b1  p:name  "Stanley Kubrick" .
	  _:b1  p:birthyear  1928 .
	  _:b2  p:name  "Orson Welles" .
	  _:b2  r:mentioned  _:b1 .
	  <<_:b1 r:influencedBy _:b2>>  p:certainty  0.9 .
	\end{verbatim}
\end{small}
	The symbols {\normalfont\texttt{\_:b1}} and {\normalfont\texttt{\_:b2}} represent RDF blank nodes in Turtle and in \turtleStar\ format. Blank nodes can be understood as data\-set-spe\-cif\-ic identifiers that cannot be referred to from other datasets~(in contrast to IRIs). Hence, a blank node in some RDF or \rdfStar\! data is similar to a vertex identity in a Property Graph. However, while this example assumes a mapping of vertices to blank nodes, the transformation also allows a user to specify a mapping of vertices to IRIs (cf.~Definition~\ref{def:VertexIdMapping}).
\end{example}

\noindent
	For this transformation, there also exists a minor limitation: 
The transformation cannot be used for a Property Graph that contains distinct edges with the same start node, the same end node, and the same label.%
	\footnote{It remains to be seen whether this limitation has an impact in practice (there exist alternative, more explicit approaches to represent information in a Property Graph without having to use multiple edges between vertices).}
Section~\ref{sec:TransformPGtoRDFstar}
	elaborates more on this limitation and formalizes the transformation.

	As a counterpart to the first two transformations, this third transformation enables RDF-based systems to import Property Graphs and execute SPARQL or \sparqlStar\ queries over the resulting RDF or \rdfStar\! data. On the other hand, a Property Graph system may provide virtual \rdfStar\ views of its Property Graphs that can be queried using \sparqlStar.

\begin{example} \label{ex:QueryingResultOf1stTransformation}
	Given the \rdfStar\ representation in Example~\ref{ex:InformalResultOf2ndTransformation}, a user can query the data originally represented in the Property Graph in Figure~\ref{fig:ExamplePG} by using \sparqlStar\ queries such as the following.
\begin{small}
	\begin{verbatim}
	    SELECT ?n WHERE {
	        ?w  p:name  "Orson Welles" .
	        <<?p r:influencedBy ?w>>  p:certainty  ?c .
	        ?p  p:name  ?n .
	    }
	    ORDER BY ?c
	\end{verbatim}
\end{small}
	The query returns an ordered list of persons influenced by Orson Welles; the order represents the certainty of whether Welles influenced these persons.
%
%
	An equivalent 
	Cypher query
		that can be evaluated over
	the original Property Graph is given as follows.
\begin{small}
	\begin{verbatim}
	    START p=node(*)
	    MATCH (p)-[r:influencedBy]->(w { name="Orson Welles" })
	    RETURN p.name
	    ORDER BY r.certainty
	\end{verbatim}
\end{small}
\end{example}

\noindent
The remainder of this document defines the proposed transformations formally. As a basis, Section~\ref{sec:Preliminaries} provides a formalization of the data models.
Thereafter, Section~\ref{sec:TransformRDFstarToRDFlikePGs} defines the lossless transformation of \rdfStar\! data to RDF-like Property Graphs~(as outlined in Section~\ref{ssec:Informal:TransformRDFstarToRDFlikePGs} above), Section~\ref{sec:TransformRDFstarToSimplePGs} defines the lossy transformation of \rdfStar\! data to simple Property Graphs~(cf.~Section~\ref{ssec:Informal:TransformRDFstarToSimplePGs} above), and Section~\ref{sec:TransformPGtoRDFstar} formalizes the transformation of Property Graphs to \rdfStar~(cf.~Section~\ref{ssec:Informal:TransformPGtoRDFstar} above).
\section{Formalization of the Data Models} \label{sec:Preliminaries}

This section recalls the relevant definitions of RDF and \rdfStar\!, provides a formal definition of the notion of a Property Graph, and identifies a class of \rdfStar\! data that
	has an expressive power analogous to Property Graphs and, thus, can be used as a basis for the transformations in this~document.

%

\subsection{\rdfStar}

\rdfStar\ is an extension of the RDF data model that makes metadata~statements a first class citizen. This section provides the relevant parts of the formalization of RDF~\cite{Cyganiak14:RDFConcepts} and \rdfStar~\cite{Hartig14:RDFStar}.

	Assume pairwise disjoint sets $\symAllIRIs$ (all IRIs), $\symAllBNodes$ (blank nodes), and $\symAllLiterals$ (literals). Each literal is associated with the IRI of a \emph{datatype}~\cite{Cyganiak14:RDFConcepts}. Hence, formally, there exists a (total) mapping $\textrm{dtype} \!: \symAllLiterals \rightarrow \symAllIRIs$. Additionally, literals may be associated with a \emph{language tag}~\cite{Cyganiak14:RDFConcepts}. Hence, there exists a partial mapping $\textrm{lang} \!: \symAllLiterals \rightarrow S_\mathsf{LTags}$ such that $S_\mathsf{LTags}$ is the set of all strings that represent well-formed language tags as defined in~\cite{Phillips09:LanguageTags}. \par An
\emph{RDF triple} is a tuple $\tuple{s,p,o} \in (\symAllIRIs \cup \symAllBNodes) \times \symAllIRIs \times (\symAllIRIs \cup \symAllBNodes \cup \symAllLiterals)$ and
	a set of RDF triples is called an \emph{RDF graph}.


\rdfStar\ extends such triples by permitting the embedding of a given triple in the subject or object position of another triple. Triples whose subject or object is an embedded triple represent some form of metadata. An embedded triple may itself be a metadata triple and, thus, may also contain embedded triples; and so forth.
%
	The following definition captures this notion.

\begin{definition} \label{def:TripleStar}
	Let $\symAllStarTriplesInfinite$ be an (infinite) set of tuples that is defined recursively as follows:
	\begin{enumerate}
		\item
			$\symAllStarTriplesInfinite$ includes all RDF triples, and
		\item
			if $t \in \symAllStarTriplesInfinite$ and $t' \in \symAllStarTriplesInfinite$, then $\tuple{t,p,o} \in \symAllStarTriplesInfinite$, $\tuple{s,p,t} \in \symAllStarTriplesInfinite$, and $\tuple{t,p,t'} \in \symAllStarTriplesInfinite$ for all $s \in (\symAllIRIs \cup \symAllBNodes)$, $p \in \symAllIRIs$, and $o \in (\symAllIRIs \cup \symAllBNodes \cup \symAllLiterals)$.
	\end{enumerate}
	Moreover, for any tuple $\tuple{s,p,o} \in \symAllStarTriplesInfinite$ and any natural number $k$, call the tuple \emph{$k$-nested} if either:
	\begin{enumerate}
		\item
			$k=0$ and the tuple $\tuple{s,p,o}$ is an RDF triple, or
		\item			
				$k>0$, for each~(embedded) tuple $t' \in \bigl( \lbrace s,o \rbrace \cap \symAllStarTriplesInfinite \bigr)$, there exists a $k'$ such that $k' \leq k - 1$ and $t'$ is $k'$-nested, and there exists an (embedded) tuple $t' \in \bigl( \lbrace s,o \rbrace \cap \symAllStarTriplesInfinite \bigr)$ that is $(k\!-\!1$)-nested.
	\end{enumerate}
	Then, an \definedTerm{\rdfStarTriple} is a tuple $t \in \symAllStarTriplesInfinite$ for which there exists a natural number $k$ such that $t$ is $k$-nested.
	A set of {\rdfStarTriple}s is called an \definedTerm{\rdfStarGraph}.
\end{definition}

\noindent
Note that the notion of \emph{$k$-nestedness}
in Definition~\ref{def:TripleStar} rules out infinitely deep nesting of {\rdfStarTriple}s.

	Hereafter, the
set of \emph{all {\rdfStarTriple}s} is denoted by
	$\symAllStarTriples$,
%
	and, for any \rdfStarTriple\ $t \in \symAllStarTriples$, $\fctTermsPlus{t}$ denotes the set of all RDF terms and all {\rdfStarTriple}s mentioned in $t$; i.e., let $t = \tuple{s,p,o}$, then $\fctTermsPlus{t} \definedAs \lbrace s,p,o \rbrace \cup \big\lbrace x' \in \fctTermsPlus{x} \,\big|\, x \in \lbrace s,o \rbrace \cap \symAllStarTriples \big\rbrace$. An \rdfStarTriple\ $t$ with $\fctTermsPlus{t} \cap \symAllStarTriples \neq \emptyset$ is called a \emph{metadata triple} (note that any other \rdfStarTriple\ is an ordinary RDF triple).



Overloading function $\fctsymTermsPlus$\!, for any \rdfStarGraph\
	$\symRDFstarGraph$\!,
$\fctTermsPlus{\symRDFstarGraph} \definedAs \bigcup_{t \in \symRDFstarGraph} \fctTermsPlus{t}$. Furthermore, $\fctTRefs{\symRDFstarGraph}$ denotes the set of all {\rdfStarTriple}s that are (recursively) embedded in {\rdfStarTriple}s of \rdfStarGraph\ $\symRDFstarGraph$; i.e., $\fctTRefs{\symRDFstarGraph} \definedAs \fctTermsPlus{\symRDFstarGraph} \cap \symAllStarTriples$.

\begin{example} \label{ex:RDFstarGraph}
	Formally,
		the \turtleStar\ document in Example~\ref{ex:RDFstarInformal} represents the following \rdfStarGraph.
	\begin{align*}
		\symRDFstarGraph_\mathsf{ex} = \big\lbrace
			& \bigtuple{ \tuple{\mathSmallRDF{ex:alice}, \mathSmallRDF{foaf:knows}, \mathSmallRDF{ex:bob}}, \, \mathSmallRDF{ex:certainty}, \, \mathSmallRDF{0.5}} , \\
			& \tuple{\mathSmallRDF{ex:alice}, \, \mathSmallRDF{foaf:name}, \, \mathSmallRDF{"Alice"}} ,\\
			& \tuple{\mathSmallRDF{ex:bob}, \, \mathSmallRDF{foaf:name}, \, \mathSmallRDF{"Bob"}} ,\\
			& \bigtuple{ \tuple{\mathSmallRDF{ex:bob}, \mathSmallRDF{foaf:age}, \mathSmallRDF{23}}, \, \mathSmallRDF{ex:certainty}, \, \mathSmallRDF{0.9}} 
		\big\rbrace
		\intertext{Note that this \rdfStarGraph\ consists of four {\rdfStarTriple}s, two of which are ordinary RDF triples and the other are metadata triples. It holds that}
		\fctTermsPlus{\symRDFstarGraph_\mathsf{ex}} = \big\lbrace
			& \mathSmallRDF{ex:alice}, \, \mathSmallRDF{foaf:knows}, \, \mathSmallRDF{ex:bob}, \tuple{\mathSmallRDF{ex:alice}, \mathSmallRDF{foaf:knows}, \mathSmallRDF{ex:bob}}, \, \mathSmallRDF{ex:certainty}, \, \mathSmallRDF{0.5}, \\
			& \mathSmallRDF{foaf:name}, \, \mathSmallRDF{"Alice"}, \, \mathSmallRDF{"Bob"}, \, \mathSmallRDF{foaf:age}, \, \mathSmallRDF{23}, \tuple{\mathSmallRDF{ex:bob}, \mathSmallRDF{foaf:age}, \mathSmallRDF{23}}, \, \mathSmallRDF{0.9} \,
		\big\rbrace \\[-3mm]
		\intertext{and, thus,} \\[-8mm]
		\fctTRefs{\symRDFstarGraph_\mathsf{ex}} = \big\lbrace
			& \tuple{\mathSmallRDF{ex:alice}, \mathSmallRDF{foaf:knows}, \mathSmallRDF{ex:bob}}, \tuple{\mathSmallRDF{ex:bob}, \mathSmallRDF{foaf:age}, \mathSmallRDF{23}}
		\big\rbrace
		\, .
	\end{align*}
\end{example}

\noindent
For a more detailed discussion of \rdfStar\!, including a well-de\-fined transformation from {\rdfStarGraph}s to ordinary RDF graphs, an \rdfStar\!-enabled extension of the Turtle syntax (as used in Examples \ref{ex:RDFstarInformal} and~\ref{ex:InformalResultOf2ndTransformation}), and a definition of \sparqlStar\ (the \rdfStar\!-aware extension of SPARQL; cf.~Examples
	\ref{ex:FirstSPARQLStarQueries} and~\ref{ex:QueryingResultOf1stTransformation}),
refer to~\cite{Hartig14:RDFStar}.

\subsection{Property Graphs}


This section defines the notion of a Property Graph formally. As a basis for this definition, assume a (programming language specific) set $\symAllDataTypes$ of data types, including a type for strings---hereafter, denoted by $\symDataTypeString$~(i.e., $\symDataTypeString \in \symAllDataTypes$). Note that $\symAllDataTypes$ may also contain data types for collections.
For each data type $\symDataType \in \symAllDataTypes$, let $\fctDom{\symDataType}$ denote the value space of $\symDataType$; i.e., the set of all possible values of type $\symDataType$. Hence, $\fctDom{\symDataTypeString}$ is the set of all strings.
Furthermore, assume that values are unique; that is, for any value $x \in \fctDom{\symDataType}$ of any data type $\symDataType \in \symAllDataTypes$, there does not exist another data type $\symDataType' \in \symAllDataTypes \setminus \lbrace \symDataType \rbrace$ such that $x \in \fctDom{\symDataType'}$.

Given such a set of data types $\symAllDataTypes$, the notion of a \emph{Property Graph} can be formalized as follows.

\begin{definition} \label{def:PG}
	Let $P$ be the (infinite) set of all possible properties; that is, a pair $p = \tuple{k,v}$ where $k \in \fctDom{\symDataTypeString}$ and $v \in \bigcup_{\symDataType \in \symAllDataTypes} \fctDom{\symDataType}$; i.e., $P = \fctDom{\symDataTypeString} \times \bigcup_{\symDataType \in \symAllDataTypes} \fctDom{\symDataType}$.
	A \definedTerm{Property Graph} is a tuple $\symPGraph = \tuple{V,E,\mathit{src},\mathit{tgt},\mathit{lbl},\mathfrak{P}}$ such that
	\begin{itemize}
		\item
			$\tuple{V,E,\mathit{src},\mathit{tgt},\mathit{lbl}}$ is an edge-labeled directed multigraph with
			\begin{itemize}
				\item a set of vertices $V$,
				\item a set of edges $E$,
				\item a function $\mathit{src} \!: E \rightarrow V$ that associates each edge with its source vertex,
				\item a function $\mathit{tgt} \!: E \rightarrow V$ that associates each edge with its target vertex, and
				\item a function $\mathit{lbl} \!: E \rightarrow \fctDom{\symDataTypeString}$ that associates each edge its label; and
			\end{itemize}
		\item
			$\mathfrak{P}$ is a function $\mathfrak{P} \!: ( V \cup E ) \rightarrow 2^P$\!.
	\end{itemize}
\end{definition}

\newpage 

\noindent
The following definition identifies a more restricted class of Property Graphs; that is, Property Graphs in which, for each vertex or edge, all properties of this vertex or edge have a unique key.

\begin{definition} \label{def:PropertyUniqueness}
	A Property Graph $\symPGraph = \tuple{V,E,\mathit{src},\mathit{tgt},\mathit{lbl},\mathfrak{P}}$ is \definedTerm{prop\-er\-ty-unique} if, for each vertex or edge $x \in ( V \cup E )$,
		$k_1 \neq k_2$ for all pairs of distinct properties $\tuple{k_1,v_1}, \tuple{k_2,v_2} \in \mathfrak{P}(x)$. 
\end{definition}

\vspace{-3mm} 

\begin{example} \label{ex:PGformal}
	Reconsider the Property Graph introduced in Example~\ref{ex:PGinformal} (cf.~Figure~\ref{fig:ExamplePG}). A formal representation of this graph is given by the tuple $\symPGraph_\mathsf{ex} = \tuple{V_\mathsf{ex},E_\mathsf{ex},\mathit{src}_\mathsf{ex},\mathit{tgt}_\mathsf{ex},\mathit{lbl}_\mathsf{ex},\mathfrak{P}_\mathsf{ex}}$ consisting of the following elements:
	\begin{itemize}
		\item $V_\mathsf{ex} = \lbrace \mathsf{Kubrick}, \mathsf{Welles} \rbrace$
		\item $E_\mathsf{ex} = \lbrace e_1, e_2 \rbrace$
		\item $\mathit{src}_\mathsf{ex}(e_1) = \mathsf{Welles}$, $\mathit{tgt}_\mathsf{ex}(e_1) = \mathsf{Kubrick}$, $\mathit{lbl}_\mathsf{ex}(e_1) = \str{mentioned}$
		\item $\mathit{src}_\mathsf{ex}(e_2) = \mathsf{Kubrick}$, $\mathit{tgt}_\mathsf{ex}(e_2) = \mathsf{Welles}$, $\mathit{lbl}_\mathsf{ex}(e_2) = \str{influencedBy}$
		\item $\mathfrak{P}_\mathsf{ex}(\mathsf{Kubrick}) = \big\lbrace \tuple{\str{name},\str{Stanley Kubrick}}, \tuple{\str{birthyear},1928} \big\rbrace$
		\item $\mathfrak{P}_\mathsf{ex}(\mathsf{Welles}) = \big\lbrace \tuple{\str{name},\str{Orson Welles}} \big\rbrace$
		\item $\mathfrak{P}_\mathsf{ex}(e_1) = \emptyset$
		\item $\mathfrak{P}_\mathsf{ex}(e_2) = \big\lbrace \tuple{\str{certainty},0.8} \big\rbrace$
	\end{itemize}
	Apparently, this example Property Graph $\symPGraph_\mathsf{ex}$ is prop\-er\-ty-unique.
\end{example}

\noindent
While the given definition
of a Property Graph does not restrict the types of property values, the remainder of this document assumes that all values of the data types in $\symAllDataTypes$ can be mapped
	to distinct
RDF literals. More precisely, the
	following definitions
assume a bijective function $vm \!: \symAllDataValues \rightarrow \symAllLiterals^\star$ such that $\symAllDataValues = \bigcup_{\symDataType \in \symAllDataTypes} \fctDom{\symDataType}$ and $\symAllLiterals^\star$ is a set of literals; i.e., $\symAllLiterals^\star \subseteq \symAllLiterals$ and $\left| \symAllLiterals^\star \right| = \left| \symAllDataValues \right|$. Hereafter,
function $vm$ is called the \emph{val\-ue-to-lit\-er\-al mapping}.
Note that this assumption rules out Property Graphs with property values that are of some data type for collections.

\subsection{Property Graph Convertibility of {\rdfStarGraph}s} \label{ssec:Preliminaries:Convertibility}

Before going into the details of the transformations it is important to note that the \rdfStar\ data model is more expressive than the Property Graphs model. For instance, \rdfStar\ allows for an arbitrarily deep nesting of metadata triples, whereas a Property Graph cannot contain additional metadata about a property of a vertex or an edge~(i.e., a property in a Property Graph cannot be annotated with properties itself). As a consequence, any transformation from {\rdfStarGraph}s to Property Graphs that adapts the natural approach of representing metadata triples as edge properties is possible only for specific {\rdfStarGraph}s. Informally, these {\rdfStarGraph}s must satisfy the following~conditions:
\begin{enumerate}
	\item \label{cnd:Convertible:Informally:1}
		Metadata triples are not nested within one another. 
	\item \label{cnd:Convertible:Informally:2}
		Metadata triples embed triples as their subject only (not as their object).
	\item \label{cnd:Convertible:Informally:3}
		The object of any metadata triple must be a literal. 
	\item \label{cnd:Invertible:Informally:5}
		For any literal in the triples it must be possible to convert the literal to a data value.
\end{enumerate}

\noindent
	The following definition formalizes these conditions.

\begin{definition} \label{def:Convertibility}
	An \rdfStarGraph\ $\symRDFstarGraph$ is \definedTerm{Property Graph convertible} (\definedTerm{PG-con\-vert\-ible} for short) if each \rdfStarTriple\ $t \in \symRDFstarGraph$ with $t = \tuple{s,p,o}$ has the following properties:
	\begin{enumerate}
		\item If $s \in \symAllStarTriples$, then $\fctTermsPlus{s} \cap \symAllStarTriples = \emptyset$. \hfill (i.e., $s$ is not a metadata triple; cf.~condition~\ref{cnd:Convertible:Informally:1} above)
		\item It holds that $o \notin \symAllStarTriples$. \hfill (i.e., $o$ is not an embedded triple; cf.~condition~\ref{cnd:Convertible:Informally:2} above)
		\item If $s \in \symAllStarTriples$, then $o \in \symAllLiterals$. \hfill (i.e., $o$ is a literal if $t$ is a metadata triple; cf.~condition~\ref{cnd:Convertible:Informally:3} above)
		\item For every literal $l \in \symAllLiterals$, if $l \in \fctTermsPlus{t}$, then $l \in \fctDom{vm^{-1}}$; where $vm^{-1}$ is the inverse of the aforementioned (bijective) val\-ue-to-lit\-er\-al mapping $vm$. \hfill (cf.~condition~\ref{cnd:Invertible:Informally:5} above)
	\end{enumerate}
\end{definition}

\begin{example}
	The sample \rdfStarGraph\ $\symRDFstarGraph_\mathsf{ex}$ (cf.~Example~\ref{ex:RDFstarGraph}) is PG-con\-vert\-ible.
\end{example}

\noindent
An even stronger notion of PG-con\-vert\-ibility is necessary for the
	second
transformation. As mentioned in Section~\ref{ssec:Informal:TransformRDFstarToSimplePGs}, this transformation can be applied only to {\rdfStarGraph}s that satisfy the conditions for PG-con\-vert\-ibility and, additionally, do not contain metadata triples about attribute triples.
The following definition captures this stronger notion of PG-con\-vert\-ibility.

\begin{definition} \label{def:StrongConvertibility}
	An \rdfStarGraph\ $\symRDFstarGraph$ is
		\definedTerm{strongly PG-con\-vert\-ible}
	if $\symRDFstarGraph$ is PG-con\-vert\-ible
		and, for every (embedded) \rdfStarTriple\ $\tuple{s,p,o} \in \fctTRefs{\symRDFstarGraph}$, it holds that $o \notin \symAllLiterals$.
\end{definition}

\begin{example} \label{ex:StrongConvertibility}
	The sample \rdfStarGraph\ $\symRDFstarGraph_\mathsf{ex}$ (cf.~Example~\ref{ex:RDFstarGraph}) is \emph{not} strongly PG-con\-vert\-ible because it contains the metadata triple $\bigtuple{ \tuple{\mathSmallRDF{ex:bob}, \mathSmallRDF{foaf:age}, \mathSmallRDF{23}}, \, \mathSmallRDF{ex:certainty}, \, \mathSmallRDF{0.9}}$.
	However, the following subgraph $\symRDFstarGraph_\mathsf{ex'} \subseteq \symRDFstarGraph_\mathsf{ex}$, which does not contain this metadata triple, is strongly PG-con\-vert\-ible.
	\begin{align*}
		\symRDFstarGraph_\mathsf{ex'} = \big\lbrace
			& \bigtuple{ \tuple{\mathSmallRDF{ex:alice}, \mathSmallRDF{foaf:knows}, \mathSmallRDF{ex:bob}}, \, \mathSmallRDF{ex:certainty}, \, \mathSmallRDF{0.5}} , \\
			& \tuple{\mathSmallRDF{ex:alice}, \, \mathSmallRDF{foaf:name}, \, \mathSmallRDF{"Alice"}} ,\\
			& \tuple{\mathSmallRDF{ex:bob}, \, \mathSmallRDF{foaf:name}, \, \mathSmallRDF{"Bob"}}
		\big\rbrace
		.
	\end{align*}	
\end{example}
\section{Transforming {\rdfStar}\! Graphs to RDF-like Property Graphs} \label{sec:TransformRDFstarToRDFlikePGs}

This section formalizes the lossless transformation of {\rdfStarGraph}s to Property Graphs as outlined in Section~\ref{ssec:Informal:TransformRDFstarToRDFlikePGs}. Recall that this transformation represents any ordinary RDF triple~(which is not a metadata triple) as an edge, and metadata triples are represented as edge properties.
Due to the aforementioned differences in the expressiveness of \rdfStar\ and Property Graphs, the transformation can be applied only to PG-con\-vert\-ible {\rdfStarGraph}s (cf.~Definition~\ref{def:Convertibility}).

Another requirement for the transformation is an unambiguous mapping of any possible IRI to a distinct string. The usual string representation of IRIs is sufficient for this purpose. Hence, the following definition assumes an injective function $im \!: \symAllIRIs \rightarrow \fctDom{\symDataTypeString}$. Hereafter, this function $im$ is called the \emph{IRI-to-string mapping}.

Given these preliminaries, the transformation is defined as follows.

\newpage 

\begin{definition}
	Let $\symRDFstarGraph$ be an \rdfStarGraph\ that is PG-con\-vert\-ible.
	Furthermore, let:
	\begin{align*}
		\mathrm{meta}(\symRDFstarGraph)
		&= \big\lbrace t \in \symRDFstarGraph \,\big|\, \fctTermsPlus{t} \cap \symAllStarTriples \neq \emptyset \big\rbrace
		, \\
		\mathrm{ord}(\symRDFstarGraph)
		&= \big( \symRDFstarGraph \cup \fctTRefs{\symRDFstarGraph} \bigr) \setminus \mathrm{meta}(\symRDFstarGraph)
		, \text{ and} \\
		\mathrm{SOTerms}^{\!+}\!(\symRDFstarGraph)
		&= \big\lbrace x \in (\symAllIRIs \cup \symAllBNodes \cup \symAllLiterals) \,\big|\, \tuple{s,p,o} \in \mathrm{ord}(\symRDFstarGraph) \text{ and } x \in \lbrace s,o \rbrace \big\rbrace
		;
	\end{align*}
	i.e.,
	$\mathrm{meta}(\symRDFstarGraph)$ is the set of all metadata triples in $\symRDFstarGraph$\!,
	$\mathrm{ord}(\symRDFstarGraph)$ is the set of all ordinary RDF triples in $\symRDFstarGraph$\!,
	and $\mathrm{SOTerms}^{\!+}\!(\symRDFstarGraph)$ is the set of all RDF terms in the subject position or object position of some ordinary triple in $\symRDFstarGraph$\!.
	The \definedTerm{RDF-like Property Graph representation} of $\symRDFstarGraph$ is the Property Graph $\symPGraph = \tuple{V,E,\mathit{src},\mathit{tgt},\mathit{lbl},\mathfrak{P}}$ that has the following properties:
	\begin{enumerate}
		\item
			The set of vertices $V$ contains $n_V = \bigl| \mathrm{SOTerms}^{\!+}\!(\symRDFstarGraph) \bigr|$ vertices, each of which represents a different RDF term
				in $\mathrm{SOTerms}^{\!+}\!(\symRDFstarGraph)$%
			%
				; i.e.,
			there exists a bijective function $v \!: \mathrm{SOTerms}^{\!+}\!(\symRDFstarGraph) \rightarrow V$ that maps each RDF term $x \in \mathrm{SOTerms}^{\!+}\!(\symRDFstarGraph)$ to a (different) vertex $v(x) \in V$.
		\item
			For each IRI $\symIRI \in \symAllIRIs$ with $\symIRI \in \mathrm{SOTerms}^{\!+}\!(\symRDFstarGraph)$, the set of properties $\mathfrak{P}\bigl( v(\symIRI) \bigr)$ of vertex $v(\symIRI) \in V$ is defined as $\mathfrak{P}\bigl( v(\symIRI) \bigr) = \big\lbrace \tuple{\str{kind},\str{IRI}} , \tuple{\str{IRI},im(\symIRI)} \big\rbrace$ where $im$ is the aforementioned IRI-to-string mapping.
		\item
			For each blank node $b \in \symAllBNodes$ with $b \in \mathrm{SOTerms}^{\!+}\!(\symRDFstarGraph)$, the set of properties $\mathfrak{P}\bigl( v(b) \bigr)$ of vertex $v(b) \in V$ is defined as $\mathfrak{P}\bigl( v(b) \bigr) = \big\lbrace \tuple{\str{kind},\str{blank node}} \big\rbrace$.
		\item
			For each literal $l \in \symAllLiterals$ with $l \in \mathrm{SOTerms}^{\!+}\!(\symRDFstarGraph)$, the set of properties $\mathfrak{P}\bigl( v(l) \bigr)$ of vertex $v(l) \in V$ is defined as
			$$\mathfrak{P}\bigl( v(l) \bigr) = \big\lbrace \tuple{\str{kind},\str{literal}} , \tuple{\str{literal},vm^{-1}(l)} , \tuple{\str{datatype},im\bigl( \textrm{dtype}(l) \bigr)} \big\rbrace \cup lang$$
			where $vm^{-1}$ is the inverse of the aforementioned (bijective) val\-ue-to-lit\-er\-al mapping $vm$, and
			$$lang = \begin{cases} \big\lbrace \tuple{\str{language},\textrm{lang}(l)} \big\rbrace & \text{if } l \in \fctDom{\textrm{lang}}, \\ \emptyset & \text{else}. \end{cases}$$
		\item
			The set of edges $E$ contains $n_E = \bigl| \mathrm{ord}(\symRDFstarGraph) \bigr|$ edges, each of which represents a different RDF triple $t \in \mathrm{ord}(\symRDFstarGraph)$. Hence, there exists a bijective function $e \!: \mathrm{ord}(\symRDFstarGraph) \rightarrow E$ that maps each $t \in \mathrm{ord}(\symRDFstarGraph)$ to a (different) edge $e(t) \in E$.
		\item
			For each triple $t \in \mathrm{ord}(\symRDFstarGraph)$ with $t = \tuple{s,p,o}$, the label of edge $e(t) \in E$ is $im(p)$, and the source and target vertex
				of $e(t)$
			is $v(s)$ and $v(o)$, respectively; i.e., $\mathit{lbl}\bigl( e(t) \bigr) = im(p)$, $\mathit{src}\bigl( e(t) \bigr) = v(s)$, and $\mathit{tgt}\bigl( e(t) \bigr) = v(o)$.
		\item
			For each triple $t \in \mathrm{ord}(\symRDFstarGraph)$, the set of properties $\mathfrak{P}\bigl( e(t) \bigr)$ of edge $e(t) \in E$ is defined as $$\mathfrak{P}\bigl( e(t) \bigr) = \bigcup_{\tuple{s,p,o} \in \mathrm{meta}(\symRDFstarGraph) \text{ such that } s=t} \big\lbrace \tuple{im(p) , vm^{-1}(o)} \big\rbrace.$$
	\end{enumerate}
\end{definition}

\newpage 

\begin{example}
	Since the sample \rdfStarGraph\ $\symRDFstarGraph_\mathsf{ex}$ (cf.~Example~\ref{ex:RDFstarGraph}) is PG-con\-vert\-ible, the given transformation can be used to convert $\symRDFstarGraph_\mathsf{ex}$ to a Property Graph. Figure~\ref{fig:ExampleResultingRDFlikePG}~(in Example~\ref{ex:InformalResultOf1stTransformation}) illustrates the resulting Property Graph $\symPGraph_\mathsf{ex}'$, which is the RDF-like Property Graph representation of $\symRDFstarGraph_\mathsf{ex}$. Formally, this Property Graph is given by the tuple $\symPGraph_\mathsf{ex}' = \tuple{V_\mathsf{ex}',E_\mathsf{ex}',\mathit{src}_\mathsf{ex}',\mathit{tgt}_\mathsf{ex}',\mathit{lbl}_\mathsf{ex}',\mathfrak{P}_\mathsf{ex}'}$ that consists of the following elements:
	\begin{itemize}
		\item $V_\mathsf{ex}' = \lbrace v_1, v_2, v_3, v_4, v_5 \rbrace$

		\item $E_\mathsf{ex}' = \lbrace e_1, e_2, e_4, e_4 \rbrace$

		\item $\mathit{src}_\mathsf{ex}'(e_1) = v_1$, $\mathit{tgt}_\mathsf{ex}'(e_1) = v_3$, and $\mathit{lbl}_\mathsf{ex}'(e_1) = \str{http://xmlns.com/foaf/0.1/name}$

		\item $\mathit{src}_\mathsf{ex}'(e_2) = v_1$, $\mathit{tgt}_\mathsf{ex}'(e_2) = v_2$, and $\mathit{lbl}_\mathsf{ex}'(e_2) = \str{http://xmlns.com/foaf/0.1/knows}$

		\item $\mathit{src}_\mathsf{ex}'(e_3) = v_2$, $\mathit{tgt}_\mathsf{ex}'(e_3) = v_4$, and $\mathit{lbl}_\mathsf{ex}'(e_3) = \str{http://xmlns.com/foaf/0.1/name}$

		\item $\mathit{src}_\mathsf{ex}'(e_4) = v_2$, $\mathit{tgt}_\mathsf{ex}'(e_4) = v_5$, and $\mathit{lbl}_\mathsf{ex}'(e_4) = \str{http://xmlns.com/foaf/0.1/age}$

		\item $\mathfrak{P}_\mathsf{ex}'(v_1) = \big\lbrace \tuple{\str{kind},\str{IRI}} , \tuple{\str{IRI},\str{http://example.org/alice}} \big\rbrace$

		\item $\mathfrak{P}_\mathsf{ex}'(v_2) = \big\lbrace \tuple{\str{kind},\str{IRI}} , \tuple{\str{IRI},\str{http://example.org/bob}} \big\rbrace$

		\item $\mathfrak{P}_\mathsf{ex}'(v_3) = \big\lbrace \tuple{\str{kind},\str{literal}} , \tuple{\str{literal},\str{Alice}} \big\rbrace$

		\item $\mathfrak{P}_\mathsf{ex}'(v_4) = \big\lbrace \tuple{\str{kind},\str{literal}} , \tuple{\str{literal},\str{Bob}} \big\rbrace$

		\item $\mathfrak{P}_\mathsf{ex}'(v_5) = \big\lbrace \tuple{\str{kind},\str{literal}} , \tuple{\str{literal},23} \big\rbrace$

		\item $\mathfrak{P}_\mathsf{ex}'(e_1) = \emptyset$

		\item $\mathfrak{P}_\mathsf{ex}'(e_2) = \big\lbrace \tuple{\str{http://example.org/certainty},0.5} \big\rbrace$

		\item $\mathfrak{P}_\mathsf{ex}'(e_3) = \emptyset$

		\item $\mathfrak{P}_\mathsf{ex}'(e_4) = \big\lbrace \tuple{\str{http://example.org/certainty},0.9} \big\rbrace$
	\end{itemize}
\end{example}

\vspace{2mm} 
\begin{remark}
	For any PG-con\-vert\-ible \rdfStarGraph\ $\symRDFstarGraph$ it holds that if $\symRDFstarGraph$ contains two~(distinct) metadata triples that differ only in their objects~(i.e., there exist $\tuple{s,p,o}, \tuple{s'\!,p'\!,o'} \in \symRDFstarGraph$ such that $s=s'$\!, $p=p'$\!, $o \neq o'$\!, $s \in \symAllStarTriples$, and $s' \in \symAllStarTriples$), then the RDF-like Property Graph representation of $\symRDFstarGraph$ is \emph{not} prop\-er\-ty-unique.
\end{remark}

\noindent
While the given transformation is lossless (i.e., resulting Property Graphs contain all information present in the original \rdfStarGraph), some use cases may have the stronger requirement that the original \rdfStarGraph\ can be reconstructed \emph{exactly} from its RDF-like Property Graph representation. Hence, for these use cases the transformation must be invertible.

To ensure an invertible transformation, any \rdfStarGraph\ to be transformed must
	not contain redundant {\rdfStarTriple}s; that is, {\rdfStarTriple}s that are embedded in metadata triples in the \rdfStarGraph\ and, additionally, appear directly (as a separate element) in the \rdfStarGraph. The following definition captures this notion of redundancy~formally.

\begin{definition}
	Let $\symRDFstarGraph$ be an \rdfStarGraph\ and let $t \in \symRDFstarGraph$ be an \rdfStarTriple\ in $\symRDFstarGraph$.
	\rdfStarTriple\ $t$ is \definedTerm{redundant} in $\symRDFstarGraph$ if there exists another (metadata) triple $t' \in \symRDFstarGraph$ such that $t \in \fctTermsPlus{t'}$.
\end{definition}

\noindent
{\rdfStarGraph}s must have the following minimality property to ensure an invertible transformation.

\newpage 

\begin{definition}
	An \rdfStarGraph\ $\symRDFstarGraph$ is \definedTerm{minimal} if there does not exist an \rdfStarTriple\ $t \in \symRDFstarGraph$ that is redundant in~$\symRDFstarGraph$\!.
\end{definition}

\begin{example}
	It is easily verified that the sample \rdfStarGraph\ $\symRDFstarGraph_\mathsf{ex}$ (cf.~Example~\ref{ex:RDFstarGraph}) is minimal.
\end{example}

\begin{remark}
	Note that every \rdfStarGraph\ can be converted into a minimal \rdfStarGraph\ by removing all redundant {\rdfStarTriple}s. Then, any unfolded RDF graph~\cite{Hartig14:RDFStar} of the resulting (minimal) \rdfStarGraph\ is equivalent to the corresponding unfolded RDF graph of the original \rdfStarGraph.
\end{remark}

%

\section{Transforming {\rdfStar}\! Graphs to Simple Property Graphs} \label{sec:TransformRDFstarToSimplePGs}

This section formalizes the
transformation of {\rdfStarGraph}s to Property Graphs as outlined in Section~\ref{ssec:Informal:TransformRDFstarToSimplePGs}. Recall that this transformation represents relationship triples (whose object is an IRI or a blank node) as edges; attribute triples (whose object is a literal) are converted into vertex properties for their subject. Metadata triples about relationship triples are converted into edge properties, and metadata triples about attribute triples cannot be converted by the transformation because the Property Graph model does not support metadata about (vertex) properties. The following definition captures this approach, which is applicable only to {\rdfStarGraph}s that are strongly PG-con\-vert\-ible (cf.~Definition~\ref{def:StrongConvertibility}).

\begin{definition}
	Let $\symRDFstarGraph$ be an \rdfStarGraph\ that is strongly PG-con\-vert\-ible.
	Furthermore,~let:
	\begin{align*}
		\mathrm{meta}(\symRDFstarGraph)
		&= \big\lbrace t \in \symRDFstarGraph \,\big|\, \fctTermsPlus{t} \cap \symAllStarTriples \neq \emptyset \big\rbrace
		, \\
		\mathrm{ord}(\symRDFstarGraph)
		&= \big( \symRDFstarGraph \cup \fctTRefs{\symRDFstarGraph} \bigr) \setminus \mathrm{meta}(\symRDFstarGraph)
		, \\
		\mathrm{ordA}(\symRDFstarGraph)
		&= \big\lbrace \tuple{s,p,o} \in \mathrm{ord}(\symRDFstarGraph) \,\big|\, o \in \symAllLiterals \big\rbrace
		, \\
		\mathrm{ordR}(\symRDFstarGraph)
		&= \big\lbrace \tuple{s,p,o} \in \mathrm{ord}(\symRDFstarGraph) \,\big|\, o \in (\symAllIRIs \cup \symAllBNodes) \big\rbrace
		, \text{ and} \\
		\mathrm{SO}^{\!+}\!(\symRDFstarGraph)
		&= \big\lbrace x \in (\symAllIRIs \cup \symAllBNodes) \,\big|\, \tuple{s,p,o} \in \mathrm{ord}(\symRDFstarGraph) \text{ and } x \in \lbrace s,o \rbrace \big\rbrace
		;
	\end{align*}
	i.e.,
	$\mathrm{meta}(\symRDFstarGraph)$ is the set of all metadata triples in $\symRDFstarGraph$\!,
	$\mathrm{ord}(\symRDFstarGraph)$ is the set of all ordinary RDF triples in $\symRDFstarGraph$\!,
	$\mathrm{ordA}(\symRDFstarGraph)$ is the set of all ordinary
		attribute triples in $\symRDFstarGraph$\!,
	$\mathrm{ordR}(\symRDFstarGraph)$ is the set of all ordinary
		relationship triples in $\symRDFstarGraph$\!,
	and $\mathrm{SO}^{\!+}\!(\symRDFstarGraph)$ is the set of all IRIs and blank nodes in the subject position or object position of some ordinary triple in $\symRDFstarGraph$\!.
	The \definedTerm{simple Property Graph representation} of $\symRDFstarGraph$ is the Property Graph $\symPGraph = \tuple{V,E,\mathit{src},\mathit{tgt},\mathit{lbl},\mathfrak{P}}$ that has the following properties:
	\begin{enumerate}
		\item
			The set of vertices $V$ contains $n_V = \bigl| \mathrm{SO}^{\!+}\!(\symRDFstarGraph) \bigr|$ vertices, each of which represents a different IRI or blank node $x \in \mathrm{SO}^{\!+}\!(\symRDFstarGraph)$. Hence, there exists a bijective function $v \!: \mathrm{SO}^{\!+}\!(\symRDFstarGraph) \rightarrow V$ that maps each $x \in \mathrm{SO}^{\!+}\!(\symRDFstarGraph)$ to a (different) vertex $v(x) \in V$.
		\item
			For each IRI $\symIRI \in \symAllIRIs$ with $\symIRI \in \mathrm{SO}^{\!+}\!(\symRDFstarGraph)$, the set of properties $\mathfrak{P}\bigl( v(\symIRI) \bigr)$ of vertex $v(\symIRI) \in V$ is defined as
			$$\mathfrak{P}\bigl( v(\symIRI) \bigr) = \big\lbrace \tuple{\str{IRI},im(\symIRI)} \big\rbrace \cup \bigcup_{\tuple{s,p,o} \in \mathrm{ordA}(\symRDFstarGraph) \text{ such that } s = \symIRI} \big\lbrace \tuple{im(p), vm^{-1}(o)} \big\rbrace,$$
			where $im$ is the aforementioned IRI-to-string mapping, and $vm^{-1}$ is the inverse of the aforementioned (bijective) val\-ue-to-lit\-er\-al mapping $vm$.
		\item
			For each blank node $b \in \symAllBNodes$ with $b \in \mathrm{SO}^{\!+}\!(\symRDFstarGraph)$, the set of properties $\mathfrak{P}\bigl( v(b) \bigr)$ of vertex $v(b) \in V$ is defined as
			$$\mathfrak{P}\bigl( v(b) \bigr) = \bigcup_{\tuple{s,p,o} \in \mathrm{ordA}(\symRDFstarGraph) \text{ such that } s = b} \big\lbrace \tuple{im(p), vm^{-1}(o)} \big\rbrace.$$
		\item
			The set of edges $E$ contains $n_E = \bigl| \mathrm{ordR}(\symRDFstarGraph) \bigr|$ edges, each of which represents a different~(relationship) triple $t \in \mathrm{ordR}(\symRDFstarGraph)$. Hence, there exists a bijective function $e \!: \mathrm{ordR}(\symRDFstarGraph) \rightarrow E$ that maps each $t \in \mathrm{ordR}(\symRDFstarGraph)$ to a (different) edge $e(t) \in E$.
		\item
			For each (relationship) triple $t \in \mathrm{ordR}(\symRDFstarGraph)$ with $t = \tuple{s,p,o}$, the label of edge $e(t) \in E$ is $im(p)$, and the  source and target vertex
			is $v(s)$ and $v(o)$, respectively; i.e., $\mathit{lbl}\bigl( e(t) \bigr) = im(p)$, $\mathit{src}\bigl( e(t) \bigr) = v(s)$, and $\mathit{tgt}\bigl( e(t) \bigr) = v(o)$.
		\item
			For each (relationship) triple $t \in \mathrm{ordR}(\symRDFstarGraph)$, the set of properties $\mathfrak{P}\bigl( e(t) \bigr)$ of edge $e(t) \in E$ is defined as
			$$\mathfrak{P}\bigl( e(t) \bigr) = \bigcup_{\tuple{s,p,o} \in \mathrm{meta}(\symRDFstarGraph) \text{ such that } s = t} \big\lbrace \tuple{im(p), vm^{-1}(o)} \big\rbrace.$$
	\end{enumerate}
\end{definition}

\vspace{-2mm} 

\begin{example}
	The \rdfStarGraph\ $\symRDFstarGraph_\mathsf{ex'}$, as introduced in Example~\ref{ex:StrongConvertibility}, is strongly PG-con\-vert\-ible and, thus, it can be converted to a simple Property Graph representation. Figure~\ref{fig:ExampleResultingSimplePG}~(in Example~\ref{ex:InformalResultOf3rdTransformation}) illustrates the resulting Property Graph $\symPGraph_\mathsf{ex}''$, which is the simple Property Graph representation of $\symRDFstarGraph_\mathsf{ex'}$. Formally, this Property Graph is given by the tuple $\symPGraph_\mathsf{ex}'' = \tuple{V_\mathsf{ex}'',E_\mathsf{ex}'',\mathit{src}_\mathsf{ex}'',\mathit{tgt}_\mathsf{ex}'',\mathit{lbl}_\mathsf{ex}'',\mathfrak{P}_\mathsf{ex}''}$ that consists of the following elements:
	\begin{itemize}
		\item $V_\mathsf{ex}'' = \lbrace v_1, v_2 \rbrace$

		\item $E_\mathsf{ex}'' = \lbrace e_1 \rbrace$

		\item $\mathit{src}_\mathsf{ex}''(e_1) = v_1$, $\mathit{tgt}_\mathsf{ex}''(e_1) = v_2$, and $\mathit{lbl}_\mathsf{ex}''(e_1) = \str{http://xmlns.com/foaf/0.1/knows}$

		\item $\mathfrak{P}_\mathsf{ex}''(v_1) = \big\lbrace \tuple{\str{IRI},\str{http://example.org/alice}}, \tuple{\str{http://xmlns.com/foaf/0.1/name},\str{Alice}} \big\rbrace$

		\item $\mathfrak{P}_\mathsf{ex}''(v_2) = \big\lbrace \tuple{\str{IRI},\str{http://example.org/bob}}, \tuple{\str{http://xmlns.com/foaf/0.1/name},\str{Bob}} \big\rbrace$

		\item $\mathfrak{P}_\mathsf{ex}''(e_1) = \big\lbrace \tuple{\str{http://example.org/certainty},0.5} \big\rbrace$
	\end{itemize}
\end{example}

\begin{remark}
	For any strongly PG-con\-vert\-ible \rdfStarGraph\ $\symRDFstarGraph$ it holds that if $\symRDFstarGraph$ contains two~(distinct) {\rdfStarTriple}s that differ only in their objects and these objects are literals~(i.e., there exist $\tuple{s,p,o}, \tuple{s'\!,p'\!,o'} \in \symRDFstarGraph \cup \fctTRefs{\symRDFstarGraph}$ such that $s=s'$\!, $p=p'$\!, $o \neq o'$\!, $o \in \symAllLiterals$, and $o' \in \symAllLiterals$), then the simple Property Graph representation of $\symRDFstarGraph$ is \emph{not} prop\-er\-ty-unique.
\end{remark}

\section{Transforming Property Graphs to {\rdfStar}\! Graphs} \label{sec:TransformPGtoRDFstar}

	This section defines the transformation of Property Graphs to {\rdfStarGraph}s as described in Section~\ref{ssec:Informal:TransformPGtoRDFstar}. 
The idea of this transformation is to represent each vertex of a given Property Graph either by a blank node or an IRI; each edge is represented as an ordinary~(non-metadata) triple whose subject and object are the blank node or IRI of the adjacent vertices and whose predicate is a IRI that denotes the label of the edge; moreover, vertex properties are also represented as ordinary triples, and edge properties are represented as metadata triples
	whose subject is the triple for the corresponding~edge.

While representing each edge~(including its label) as a single triple is perhaps the most intuitive approach to transform such edges, this approach has the following shortcoming: If there are two~(or more) distinct edges that connect the same vertices and have the same label~(but may have different properties), the approach would represent both edges by a single
	triple. As a result, this triple would not represent any one of the edges unambiguously when embedded in a metadata triple for a property of the edge.
To avoid this problem, the transformation is restricted to Property Graphs that do \emph{not} contain distinct edges with the same source vertex, the same target vertex, and the same label. Hereafter, these Property Graphs are called \emph{edge-unique}.

\begin{definition}
	A Property Graph $\symPGraph = \tuple{V,E,\mathit{src},\mathit{tgt},\mathit{lbl},\mathfrak{P}}$ is \definedTerm{edge-unique}
		if there does not exist a pair of edges $(e,e') \in E \times E$ such that $e \neq e'$\!, $\mathit{src}(e) = \mathit{src}(e')$, $\mathit{tgt}(e) = \mathit{tgt}(e')$, and $\mathit{lbl}(e) = \mathit{lbl}(e')$.
\end{definition}

\begin{example}
	The Property Graph $\symPGraph_\mathsf{ex}$
		in Example~\ref{ex:PGformal} is~edge-unique (cf.~Figure~\ref{fig:ExamplePG}).
\end{example}

\begin{remark}
	Note that the requirement of edge-uniqueness does not present a restriction on the expressiveness of Property Graphs. Information represented by introducing multiple edges between vertices can also be modeled by using an alternative, more explicit approach. For instance, the relationship captured by each of the multiple edges may be modeled as a separate vertex.
\end{remark}

\noindent
The transformation assumes
	three user-specified templates for generating IRIs. The first two of these templates can be used to generate
IRIs that denote arbitrary edge labels and arbitrary properties keys, respectively. The following two mappings capture the notion of these templates~formally.

\begin{definition}
	An \definedTerm{edge label mapping} $lm$ is a bijective function $lm \!: \fctDom{\symDataTypeString} \rightarrow I$ such that $I$ is a set of IRIs; i.e., $I \subseteq \symAllIRIs$  and $\left| I \right| = \left| \fctDom{\symDataTypeString} \right|$.
	\hfill (Recall that $\fctDom{\symDataTypeString}$ is the set of all strings.)
\end{definition}

\begin{definition}
	A \definedTerm{property key mapping} $km$ is a bijective function $km \!: \fctDom{\symDataTypeString} \rightarrow I$ such that~$I$ is a set of IRIs; i.e., $I \subseteq \symAllIRIs$  and $\left| I \right| = \left| \fctDom{\symDataTypeString} \right|$.
	\hfill (Recall that $\fctDom{\symDataTypeString}$ is the set of all strings.)
\end{definition}

\begin{example}
An example of a property key mapping is a function $km_\mathsf{ex}$ such that, for every string $s \in \fctDom{\symDataTypeString}$, function $km_\mathsf{ex}$ returns the IRI \texttt{http://example.org/property/}\textsf{\normalfont urlenc($s$)} where \textsf{\normalfont urlenc($s$)} is the URL encoded version of string $s$.
Similarly, $lm_\mathsf{ex}$ is an edge label mapping such that, for every string $s \in \fctDom{\symDataTypeString}$, $lm_\mathsf{ex}$ returns the IRI \texttt{http://example.org/relationship/}\textsf{\normalfont urlenc($s$)}.
\end{example}

\begin{remark}
	While a user-specified edge label mapping (resp.~a property key mapping) can be used for transforming different Property Graphs, this should be done only if the same labels (resp.~the same property keys) in these different Property Graphs have the same meaning. If that is not the case, a different edge label mapping (resp.~property key mapping) should be used.
\end{remark}

\noindent
The third user-specified template can be used to generate IRIs or blank nodes for the vertices of a Property Graph. Formally, the following mapping captures such a template.

\begin{definition} \label{def:VertexIdMapping}
	Let $\symPGraph = \tuple{V,E,\mathit{src},\mathit{tgt},\mathit{lbl},\mathfrak{P}}$ be a Property Graph.
	A \definedTerm{vertex identity mapping} $id$ for $\symPGraph$ is an injective function $id \!: V \rightarrow (\symAllBNodes \cup \symAllIRIs)$.
\end{definition}

\noindent
The transformation can now be formalized as follows.

\newpage 

\begin{definition}
Let $\symPGraph = \tuple{V,E,\mathit{src},\mathit{tgt},\mathit{lbl},\mathfrak{P}}$ be a Property Graph that is prop\-er\-ty-unique (cf.~Definition~\ref{def:PropertyUniqueness}) and edge-unique,
let $id$ be a vertex identity mapping for $\symPGraph$,
let $lm$ be an edge label mapping,
and, for each edge $e \in E$, let $\mathrm{t}_e$ be the RDF triple $\bigtuple{ id\bigl(\mathit{src}(e)\bigr), lm\bigl(\mathit{lbl}(e)\bigr), id\bigl(\mathit{tgt}(e)\bigr) }$.
Then, given a property key mapping $km$ (and the val\-ue-to-lit\-er\-al mapping $vm$), the $(id,lm,km)$-spe\-cif\-ic \definedTerm{\rdfStar\! representation} of $\symPGraph$ is the set of {\rdfStarTriple}s $\symRDFstarGraph\! \definedAs \symRDFstarGraph_\mathsf{vp} \cup \symRDFstarGraph_\mathsf{ep} \cup \symRDFstarGraph_\mathsf{en}$ that consists of the following three subsets:
\begin{description}
	\item
		$\symRDFstarGraph_\mathsf{vp} \definedAs \big\lbrace\tuple{ id(v), km(k), vm(x) } \,\big|\, v \in V \text{ and } \tuple{k,x} \in \mathfrak{P}(v) \big\rbrace$
		\hfill
		(vertex properties)
	\item
		$\symRDFstarGraph_\mathsf{ep} \definedAs \big\lbrace\tuple{ \mathrm{t}_e, km(k), vm(x) } \,\big|\, e \in E \text{ and } \tuple{k,x} \in \mathfrak{P}(e) \big\rbrace$
		\hfill
		(edges with properties)
	\item
		$\symRDFstarGraph_\mathsf{en} \definedAs \big\lbrace \mathrm{t}_e \,\big|\, e \in E \text{ and } \mathfrak{P}(e) = \emptyset \big\rbrace$
		\hfill
		(edges without properties)
\end{description}
\end{definition}

\vspace{-2mm} 

\begin{example}
	Recall the sample Property Graph $\symPGraph_\mathsf{ex}$ as given in Example~\ref{ex:PGformal} (and illustrated in Figure~\ref{fig:ExamplePG}). Assume a vertex identity mapping $id_\mathsf{ex}$ that maps each vertex in $\symPGraph_\mathsf{ex}$ to a distinct blank node; i.e.,
	$id_\mathsf{ex}( \mathsf{Kubrick} ) = b_\mathsf{K}$ and
	$id_\mathsf{ex}( \mathsf{Welles} ) = b_\mathsf{W}$, with $b_\mathsf{K}, b_\mathsf{W} \in \symAllBNodes$.
	Then, the $(id_\mathsf{ex},lm_\mathsf{ex},km_\mathsf{ex})$-spe\-cif\-ic \rdfStar\! representation of Property Graph $\symPGraph_\mathsf{ex}$ is the following \rdfStarGraph~(Example~\ref{ex:InformalResultOf2ndTransformation} provides a \turtleStar\ serialization of this \rdfStarGraph):
	\begin{align*}
		\big\lbrace \,
			& \bigtuple{b_\mathsf{K}, \text{\normalfont\texttt{http://example.org/property/name}}, \str{Stanley Kubrick}} , \\
			& \bigtuple{b_\mathsf{K}, \text{\normalfont\texttt{http://example.org/property/birthyear}}, 1928} , \\
			& \bigtuple{b_\mathsf{W}, \text{\normalfont\texttt{http://example.org/property/name}}, \str{Orson Welles}} , \\
			& \bigtuple{b_\mathsf{W}, \text{\normalfont\texttt{http://example.org/relationship/mentioned}}, b_\mathsf{K}} , \\
			& \bigtuple{ \tuple{b_\mathsf{K},\text{\normalfont\texttt{http://example.org/relationship/influencedBy}},b_\mathsf{W}},\\
			& \hspace{50mm} \text{\normalfont\texttt{http://example.org/property/certainty}}, 0.8 }
		\, \big\rbrace
		\, .
	\end{align*}
\end{example}

\section{Acknowledgements}
	The author would like to thank a number of
people for providing feedback on the proposal, as well as on earlier versions of this document. In alphabetical order, these are: Bryan Thompson, Cosmin Basca, Grant Weddell, Juan Sequeda, Kavitha Srinivas, Maria-Esther Vidal, Mike Personick, Peter Boncz, and Tamer \"Ozsu.

\bibliographystyle{alpha}
\bibliography{main}

\end{document}